\documentclass[english,groupedaddress,superscriptaddress,twocolumn]{revtex4-1}

\usepackage{graphicx,epsfig,units}
\usepackage{amsmath,amsfonts,mathrsfs,amsbsy,bm,babel}

\newcommand{\DF}{$D_\mathrm{F}$}

\begin{document}

\title{Bulk-boundary correspondence in soft matter}

\author{Mehmet Ramazanoglu}
\email{ramazanoglum@itu.edu.tr}
\affiliation{Physics Engineering Department, Istanbul Technical University, 34469, Maslak, Istanbul, Turkey}
\affiliation{Brockhouse Institute for Materials Research, Hamilton, ON L8S 4M1, Canada}

\author{\c{S}ener \"{O}z\"{o}nder}
\email{ozonder@uw.edu}
\affiliation{Physics Engineering Department, Istanbul Technical University, 34469, Maslak, Istanbul, Turkey}

\author{Rumeysa Salc{\i}  \vspace{2mm}}
\affiliation{Physics Engineering Department, Istanbul Technical University, 34469, Maslak, Istanbul, Turkey}

\begin{abstract}
Bulk-boundary correspondence is the emergence of features at the boundary of a material that are dependent on and yet distinct from the properties of the bulk of the material. The diverse applications of this idea in topological insulators as well as high energy physics prove its universality. However, whether a form of bulk-boundary correspondence holds also in soft matter such as gels, polymers, lipids and other biomaterials is thus far unknown. Aerosil-dispersed liquid crystal gels (LC+aerosil) provide a good testing ground to explore the relation between the controlled variations of the aerosil density within the liquid crystal host bulk and the surface topography of the sample. Here we report on one of the earliest if not the first direct observation of such a correspondence where 
the controlled strength of random disorder created by aerosil dispersion in the bulk liquid crystal is correlated with the fractal dimension of the surface. 
We obtained the surface topography of our gel samples with different quenched random disorder strengths by using atomic force microscope techniques, and computed the fractal dimension for each sample.
We found that an increase of the aerosil gel density in the bulk corresponds to an increase in the fractal dimension at the surface. From our results emerges a new method to acquire the bulk properties of soft matter such as density, randomness and phase merely from the fractal dimension of the surface.
\end{abstract}

%\pacs {5.25.-j, 61.05.fg}
\date{\today}
\maketitle

The connection between a material's bulk and its boundary has been one of the guiding principles in several branches of physics in the last decade. The main idea is that the boundary of the system would feature excitations that do not occur in the bulk, yet the physics on the boundary is still determined by the properties of the bulk. For example in topological insulators, the index theorem relates the Chern number quantifying the topology of the insulating bulk to the spectrum of the edge states at the boundary \cite{HasanKane,TopIandSC,KANE20133,LeeTony}. %(1206.4410)
The holographic principle in high energy physics, also known as gauge/gravity duality, is another example of the bulk-boundary correspondence where the spectrum of the strongly interacting gauge theory in four spacetime dimensions is connected to the weakly interacting theory on the three dimensional boundary via duality \cite{liu2012black,Bousso2002}.

\begin{figure}[t!]
\includegraphics[scale=0.2]{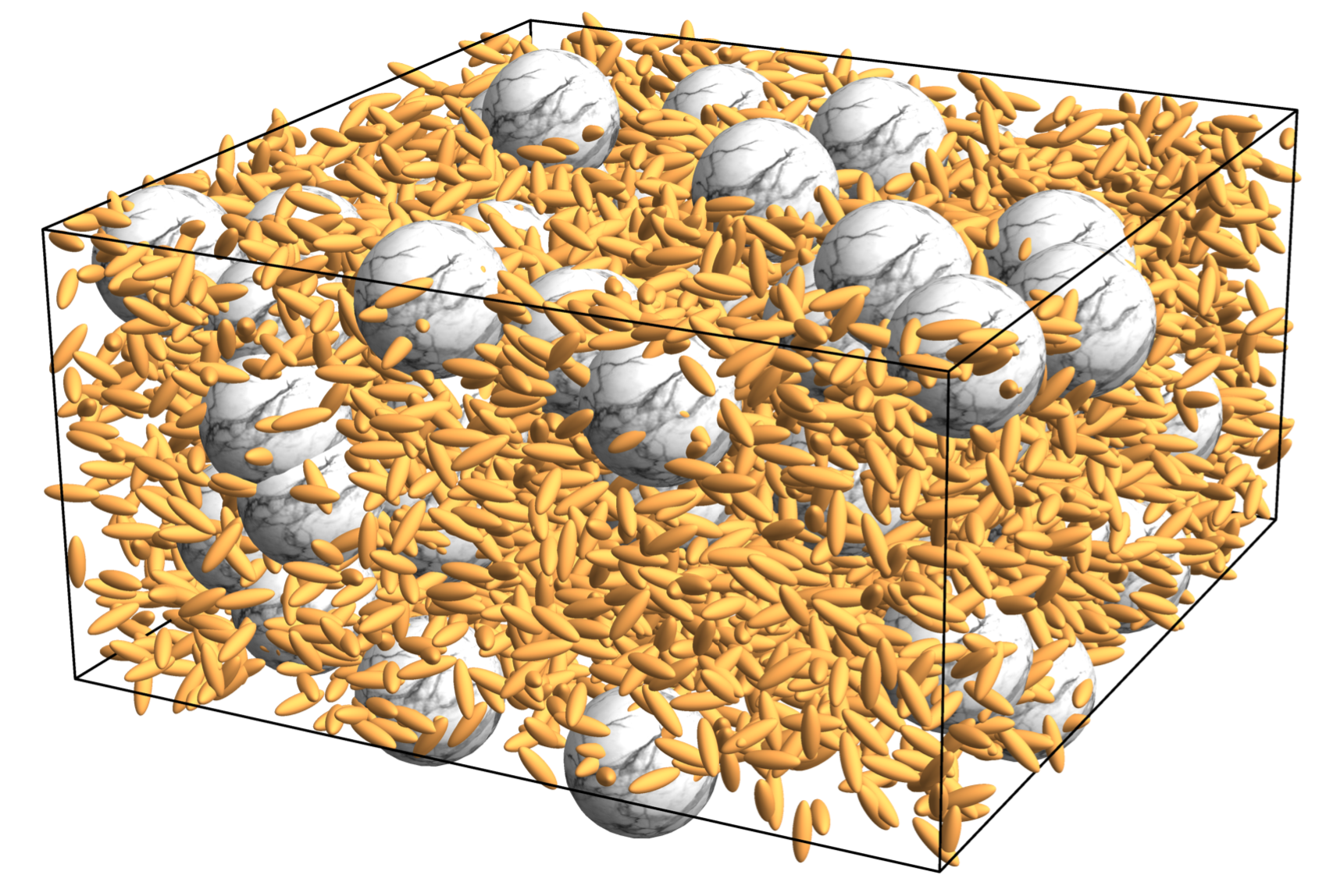}
\includegraphics[scale=0.28]{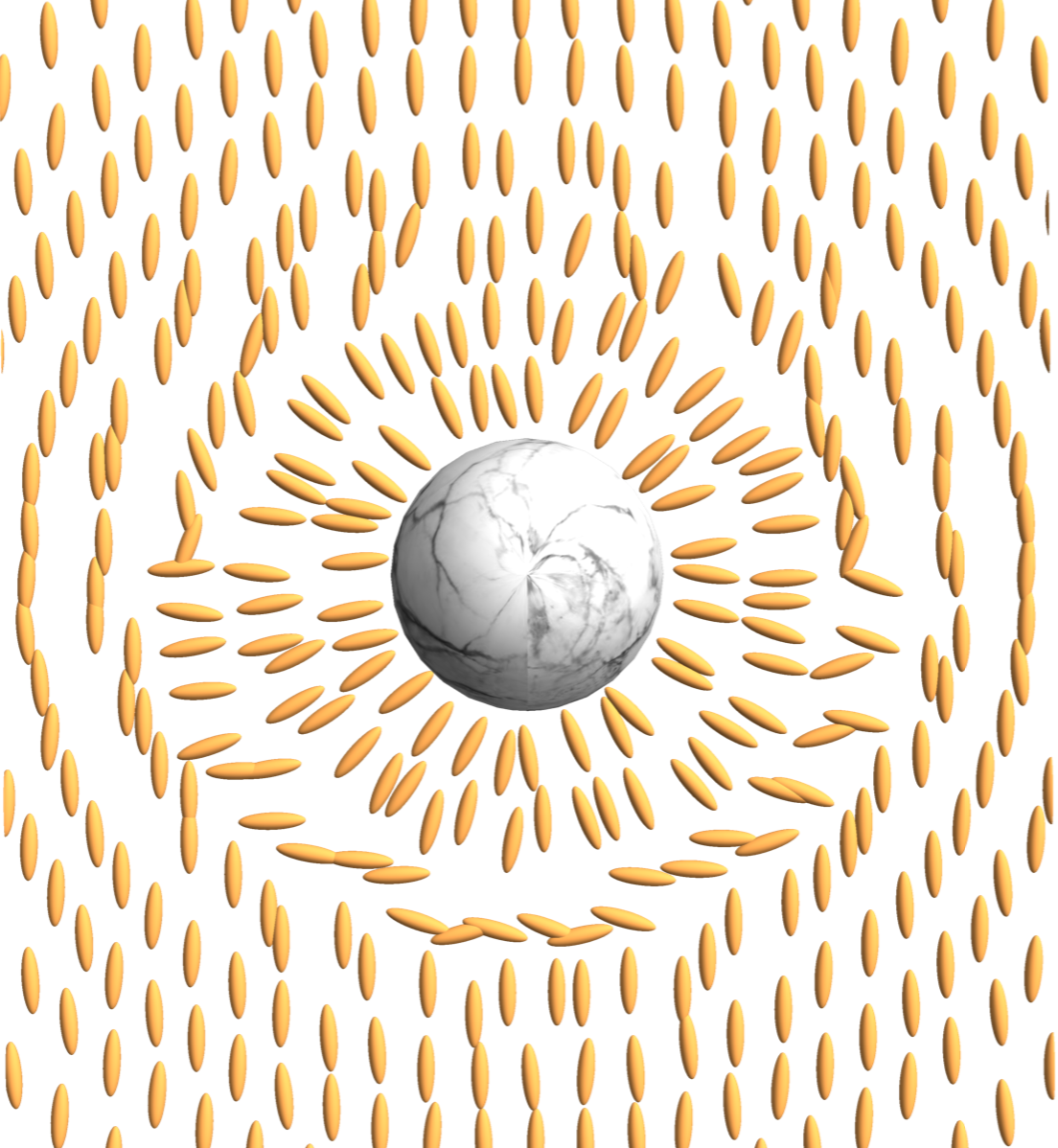}

\caption{lkjf
\textbf{Aerosil and liquid crystal (LC) mixture.}
(top) A 3D cartoon of 
dispersed aerosil nanoparticles in LCs. 
(bottom) The hydroxyl groups covering the surface of aerosils, which are {7~nm} in diameter, 
electrostatically interact with 8CB LC molecules. 
This disturbs the nematic or smectic order. 
Since the aerosil nanoparticles form a random network, the LC phase perturbation becomes random as well.
}
\label{fig1}
\end{figure}

Here we report on the first ever test of whether bulk-boundary correspondence holds in soft condensed matter systems \cite{TheOxfordHandbookofSoftCondensedMatter,kleman2007soft}, particularly in aerosil dispersed liquid crystals (Fig.~\ref{fig1}a). We prepared liquid crystal+aerosil (LC+aerosil) gel mixtures with varying amount of aerosil within, and observed that the aerosil gel density $\rho_s=m_\mathrm{SiO_2}/V_\mathrm{LC}$ in the bulk is correlated with the fractal dimension of the surface. This experimental verification of the bulk-boundary correspondence in soft matter is the main goal of this study. 

Liquid crystals (LCs)
are not only utilized in 
screens and TVs, but also used to study phase transitions \cite{deGennes}. Since they possess a rich spectrum of different phases with different types of phase transitions, they stand out as particular model systems where not only structural but also magnetic phase transitions can be imitated. Pure and doped LCs can mimic complex many-body systems like 
quantum magnets in random background fields. 
Having two order parameters, amplitude $\psi$ and phase $\alpha$, make LCs unique materials   
that closely resemble superconductors \cite{deGennes}. 
There is a strong mapping between different LC phases and the phases of the nematic spin order of the anti-ferromagnetic modulations in Fe-As based metals and 
superconductors. Thus, LCs are used to simulate and better understand certain solid state materials \cite{Zhang}. 

\begin{figure*}
\includegraphics[scale=0.34,trim={0 0.2cm 0 0},clip]{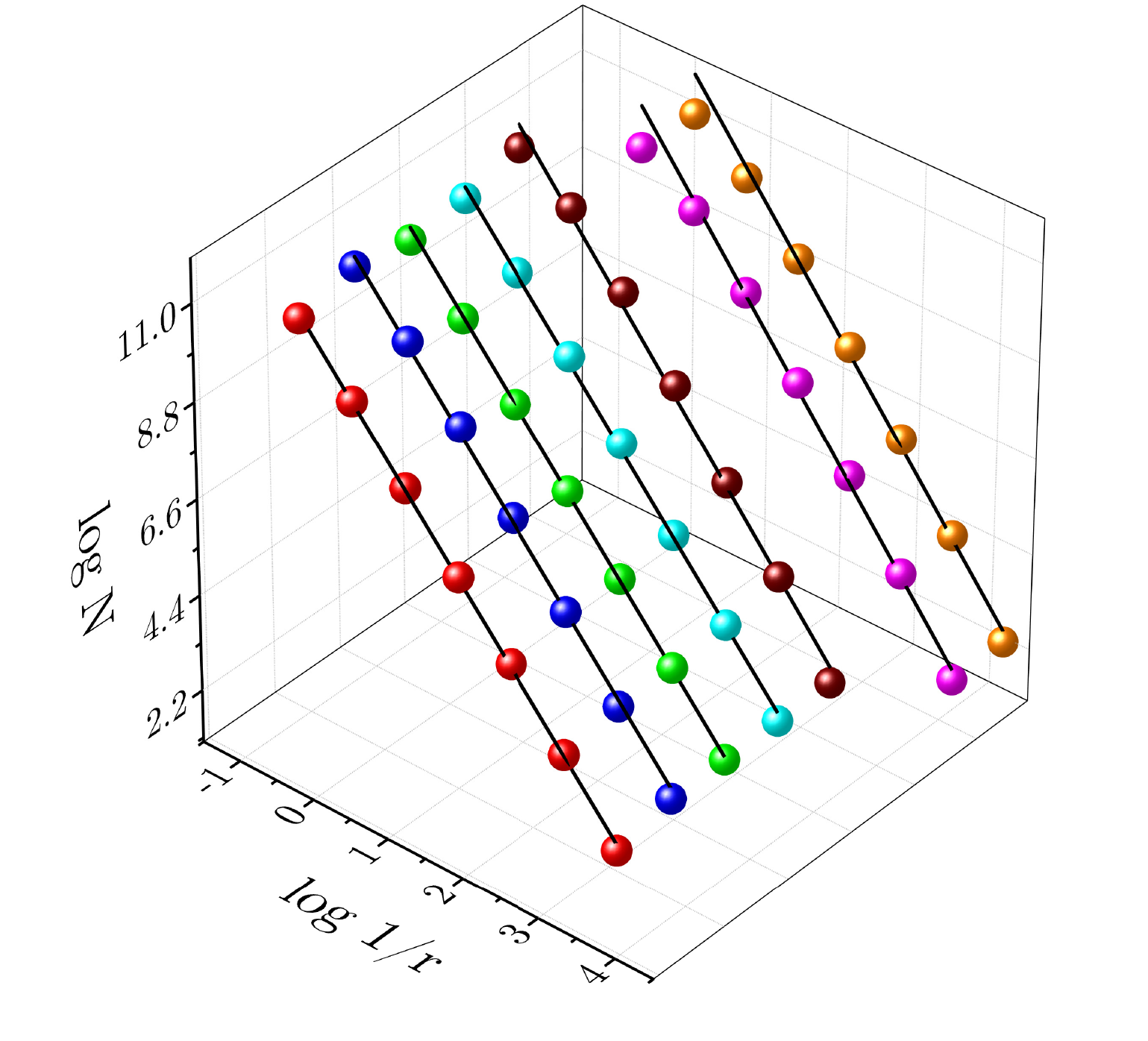}
\includegraphics[scale=0.38,trim={0 0.7cm 0 0},clip]{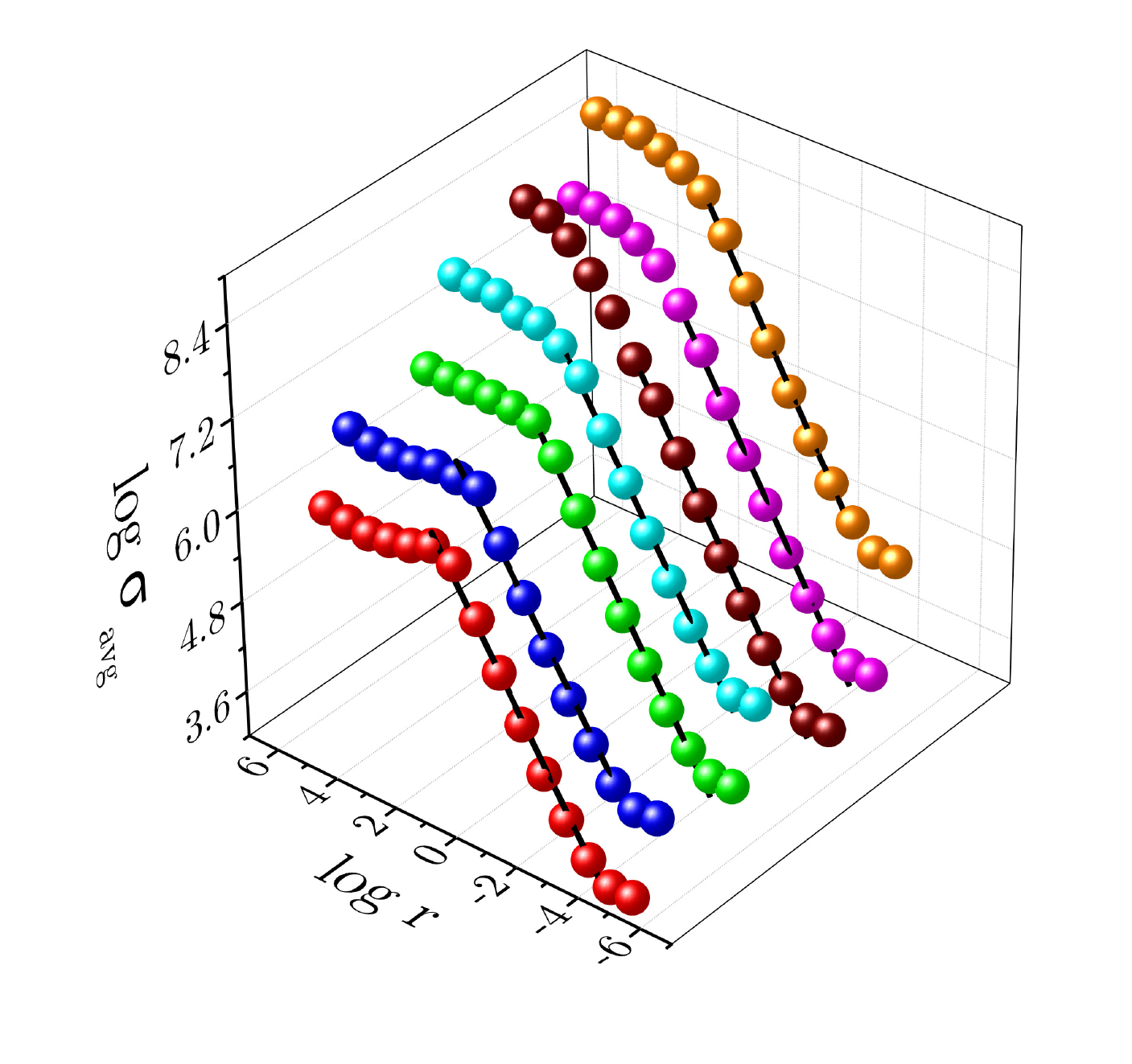}
\includegraphics[scale=0.35]{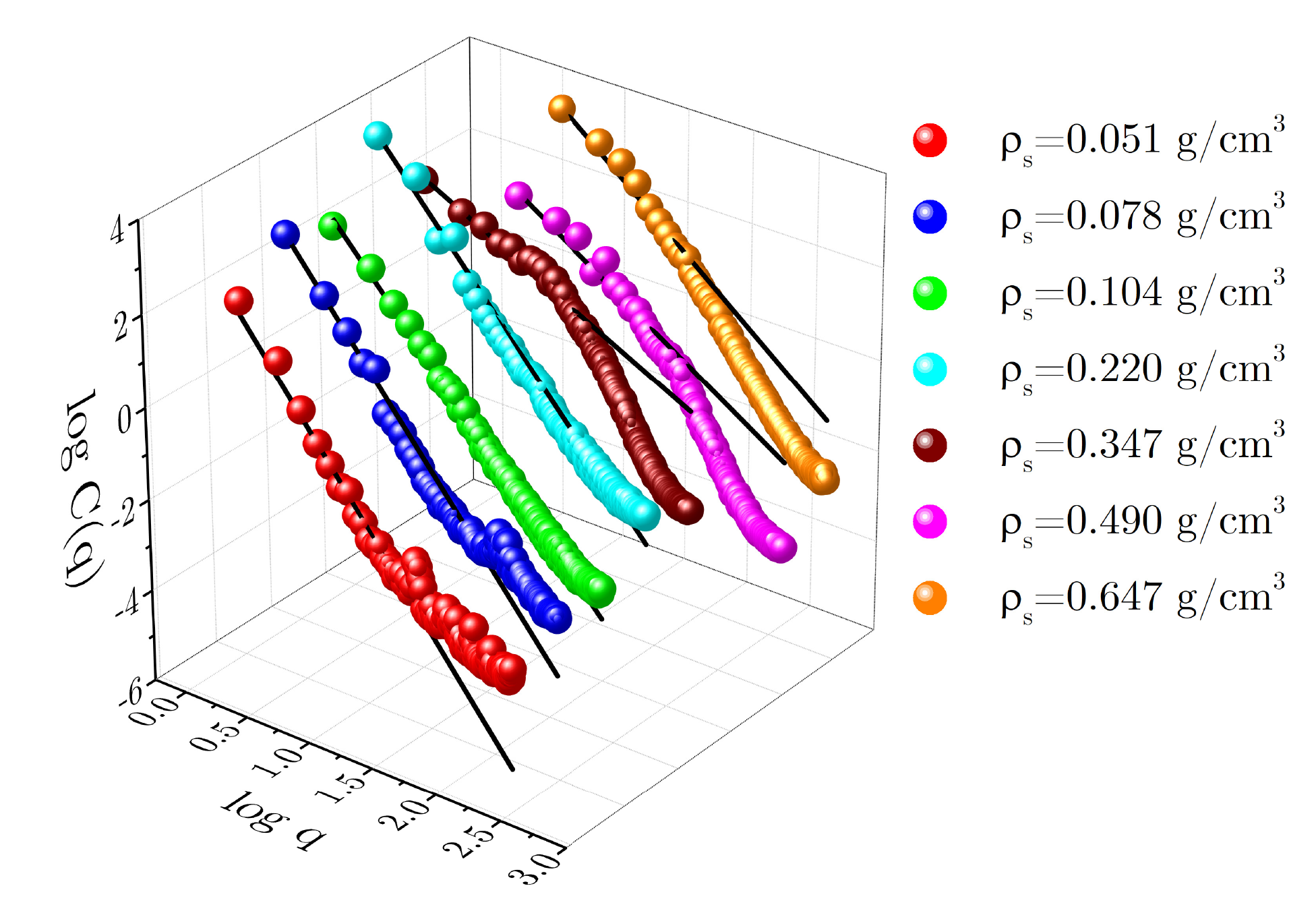}
\caption{
\textbf{Fractal dimension 
analysis of LC+aerosil gel surfaces.} (left) Box counting, (middle) coarse graining and (right) 
power spectral density calculations and 
associated  
power-law fits are shown in log-log plots. 
There are seven sets of data within each plot which correspond to our seven samples with different aerosil gel densities $\rho_s$.
}
\label{fig2}
\end{figure*}

Aerosil nanoparticles dispersed 
in LC hosts
lead to a random network of locally pinned LC molecules, thus they can be used to study the controlled random disorder effects 
and the associated phase transitions \cite{Zhou,Haga,Iannacchione,Sungil,Bellini,Leheny,Germano,mmt,mmt1,mmt2,Freelon,Garland,BelliniScience}. 
The electrostatic interaction between the aerosil surface and the polar end of the LC molecules creates pinning forces that perturb the order in the nematic and smectic LC layers (Fig.~\ref{fig1}b). Since the position of the aerosil nanoparticles are random in the LC bulk, the resultant random perturbation of LC order become quenched, a situation which is known as quenched random 
disorder \cite{Bellini,Garland,Haga}. 

With heat-capacity measurements and x-ray scattering experiments, the quenched random disorder effects on the phase transition characteristics for several LCs were studied within nematic and smectic phases.  

Aerosil dispersion within LCs not only creates quenched random disorder in the bulk, but also causes topographical changes on the material's surface.   
For example, increasing aerosil gel density within the sample makes its surface rougher \cite{Hauser,Salci}. 
Physical systems where the structural randomness can be controlled, here by varying the aerosil gel density $\rho_s$, also appear in solid state physics. One example that has been examined in random field experiments is MnZnF$_2$. 
MnZnF$_2$ is a diluted quantum anti-ferromagnet where its magnetic properties can be explained by the 3D random field Ising model. Here the long-range order of the anti-ferromagnetic interactions is disrupted by the doped non-magnetic Zn ions, and the remaining short-range anti-ferromagnetic phase can be controlled by an external field \cite{birgeneau}. 

Random pinning of aerosil particles in the bulk changes the surface while they introduce short-range modulations in the nematic and smectic phases in the bulk structure \cite{Hauser,Salci}. Therefore the randomness in the bulk reflects itself as a self-affine, fractal surface on the boundary.
In order to test this relation between the bulk and surface, we prepared seven LC+aerosil (8CB+SiO$_2$) gel mixtures with $\rho_s$=0.051, 0.078, 0.104, 0.22, 0.347, 0.49 and 0.647 g/cm$^3$. Here 8CB is the abbreviation for octylcyanobiphenyl LC.  
Since aerosil nanoparticles disrupt the ordered state of LCs, the strength of quenched random disorder in the bulk of 8CB LCs can be controlled by adjusting the aerosil gel density $\rho_s$. This parameter has also been used to characterize the LC+aerosil gel samples  \cite{Zhou,Haga,Iannacchione,Sungil,Bellini,Leheny,Germano,mmt,mmt1,mmt2,Freelon,Garland,Salci}. 

We scanned the surfaces of our samples with an atomic force microscope (AFM) and obtained surface height profiles with nanometer precision into $256{\times}256$ matrices \cite{Salci}. From the surface data, we first compute the surface fractal dimension \DF~of each sample and then investigate if \DF's are correlated with $\rho_s$'s, which are known for each sample.

\subsection*{Analysis}
Fractal surfaces appear in nature in diverse areas such as fracture in rocks, cancer growth processes, wetting of surfaces, burning of paper and rupturing \cite{Persson,barabasi}. Fractal surfaces possess self-affine structure, i.e., when observed at different scales the surface looks self-similar. This scale-independent property is 
characterized by fractal dimension \DF, which is calculated from the variation of the surface roughness with respect to the spatial scale. We use three separate methods to calculate fractal dimension of our samples' surfaces; power spectral density, box counting and coarse graining (also called variable band width) (Fig.~\ref{fig2}a-c). Since the value of the fractal dimension is dependent on the method used, each method has its own definition of fractal dimension and its own systematic and random errors. In this work, we are interested in the trend of how \DF~changes with varying aerosil gel density $\rho_s$ rather than the actual values of the fractal dimensions of  each sample.

Before calculating \DF's, we created, with simulation, mock fractal surfaces with known \DF's in order to test our codes, and from these simulations we determined systematic and random errors of the three methods we used to calculate \DF's. Then we used our codes on the real AFM surface data from our samples and our findings established that our fractal dimension analysis data from all three \DF~calculation methods fall on a straight line on a log-log plot (Fig.~\ref{fig2}a-c). This power-law feature indicates the existence of self-affine surface structure and makes the use of fractional dimension analysis justified. 

One of the methods we used to calculate \DF's of the surfaces is power spectral density (PSD). We calculated the periodograms, i.e., Fourier transform squared, for each surface data. We verified that periodograms of our data fall as power law in inverse length scale, which is characteristic of a self-affine surface. The slope of the fall off in the Fourier space gives \DF$^\mathrm{PSD}$~of that surface. When we plot \DF~versus $\rho_s$, we find a rising trend where \DF~increases with increasing aerosil density $\rho_s$ (Fig.~\ref{fig2}a). We find a similar trend by using the box counting method (Fig.~\ref{fig2}b). For this method, we first turned our surface height data from $256{\times}256$ matrices into 3D arrays of $256{\times}256{\times}256$ by rescaling the surface heights between 1 and 256. This gives us 3D arrays that contain ``1''s at coordinates that correspond to the surface heights and ``0''s for the remaining entries. Then we divide these 3D arrays into boxes of size ${r=2}$, 4, 8, 16, 32, 64 and 128, and count the number of boxes containing ``1''s for each $r$, which we call $n(r)$. We find box counting dimension \DF$^\mathrm{box}$ by fitting our counts to the curve $n(r)=r^{-D_\mathrm{F}^\mathrm{box}}$. 
The third method we employed was the coarse graining (CG) method where we first turn the $256{\times}256$ surface data into a 1D list by flattening it. Then we divide this list into equal partitions of size $r$. For a chosen $r$, we calculate the standard deviation $\sigma$ of the heights within each partition, and calculate the mean of those $\sigma$'s to find $\sigma_\mathrm{avg}$. We repeat this calculation for different values of partition size $r$, which are different powers of 2. The coarse graining fractal dimension \DF$^\mathrm{CG}$ is found by fitting the curve $\sigma_\mathrm{avg}(r)=r^{\zeta}$  to the calculated $\sigma_\mathrm{avg}$ values, and then using the definition {$D_\mathrm{F}^\mathrm{CG}=2 + \zeta$}. The coarse graining method yields an increasing trend similar to the other two methods (Fig.~\ref{fig2}c). Here we emphasize that we have made our fits to extract \DF's for each sample by using the same range of data points within all three methods, otherwise one can find any trend by arbitrarily choosing the fit range (see Supplementary Fig.~1 for the detailed plots of all the fits).

\begin{table}
\caption {\textbf{Aerosil gel densities and fractal dimensions of the samples.}
Aerosil gel density $\rho_s$, weight percentage $wt\%$, fractal dimensions obtained from the box counting, coarse graining and 
power spectral density techniques and the order parameter $\beta$ are given. Here the aerosil gel density $\rho_s$ is given by $\rho_s=m_\mathrm{aerosil}/V_\mathrm{LC}$ in units of g/cm$^3$, and it also quantifies the strength of disorder in the bulk.  
The weight percentage $wt\%$ can likewise be used to quantify the disorder strength. The values of $\beta$ are known from x-ray scattering experiments \cite{Sungil,Leheny}.  
\vspace{1mm}
}

\label{tablo1}  
\begin{tabular}{|c|c|c|c|c|c|}
\hline
 
~$\rho_s$~&~$wt\%$~&~$D_\mathrm{F}^\mathrm{box}$~ & ~$D_\mathrm{F}^\mathrm{CG}$~&~$ D_\mathrm{F}^\mathrm{PSD}$~& ~$\beta$~\\  
 \hline
 \hline
~0.051~ &  ~4.9~     &~2.02(2)~   &~0.59(5)~    &~2.12(4)~    & ~0.22(2)~ \\ 
~0.078~ &  ~7.3~     &~1.99(5)~   &~0.66(4)~    &~2.24(3)~    & ~0.23(2)~ \\
~0.104~ &  ~9.5~     &~2.04(2)~   &~0.69(2)~    &~2.44(4)~    & ~0.26(2)~ \\ 
~0.22~  &  ~18~      &~2.10(4)~   &~0.70(2)~    &~2.39(8)~    & ~0.31(2)~ \\
~0.347~ &  ~26~      &~2.29(6)~   &~0.72(2)~    &~2.72(3)~    & ~0.28(4)~ \\
~0.49~  &  ~33~      &~2.25(8)~   &~0.73(2)~    &~2.88(3)~    & ~0.39(3)~ \\
~0.647~ &  ~39.5~    &~2.23(8)~   &~0.75(1)~    &~2.85(5)~    & ~0.41(3)~ \\ 
\hline 
\end{tabular}\\ 
\vspace{1mm}
\end{table}

\begin{figure}[!h]
\includegraphics[scale=0.92]{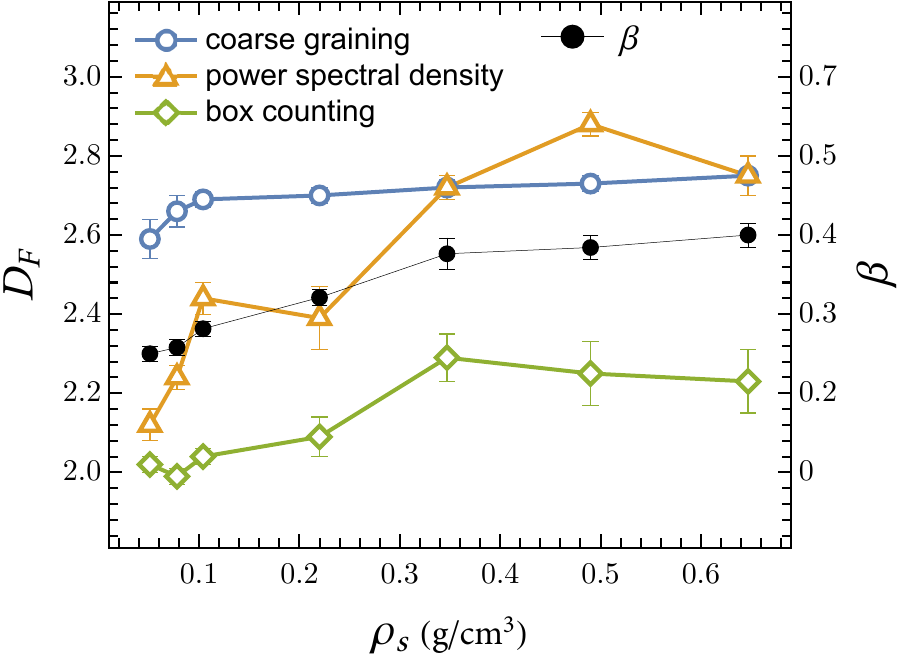}
\caption{
\textbf{Correlation between the aerosil gel density, the fractal dimension and the order parameter $\beta$.}
Calculated fractal dimensions with respect to aerosil gel density $\rho_s$, which controls the disorder strength in the bulk.
This graph also contains the order parameter $\beta$ values obtained from previous x-ray experiments \cite{Sungil,Germano,mmt1}.
We added ``2'' to the order parameter $\beta$ to be able to show all of the results on the same graph. The error bars are coming from the Marquardt-Levenberg nonlinear fit analyses.
}
\label{fig3}
\end{figure}

The results of this study (Table~\ref{tablo1}) establish a positive correlation between the bulk aerosil gel density $\rho_s$ and the surface fractal dimension \DF. 
Fig~\ref{fig3} shows that the surface fractal dimension, calculated via three different methods, increases with increasing $\rho_s$.

Since $\rho_s$ also controls the strength of the randomness in the bulk, the bulk-boundary correspondence we discovered also reveals a link between the surface fractality and the order parameter $\beta$ measured via x-ray scattering \cite{Germano,Sungil}. In Fig.~\ref{fig3}, in addition to \DF's, we also show the trend of the order parameter $\beta$ versus the aerosil gel density $\rho_s$, which can also be seen as the disorder strength. 
According to previous x-ray studies, $\beta$ exhibits an increasing trend 
with respect to increasing $\rho_s$. 
The smaller $\beta$ values fall into the universality class of the 2D Ising model. As $\rho_s$ increases, the increasing $\beta$ values go past the tricritical point and become equal to that of the universality class of the 3D Ising model.
\cite{Sungil, Garland, mmt2}. Such an increase in $\beta$ is a sign of change in the universality class of the 8CB LC's smectic-$A$ phase under the quenched random disorder effects. Our  findings (Fig.~\ref{fig3}) uncover the correlation between the order parameter $\beta$ and the fractal dimension \DF.  

Our results are also in good agreement with the 3D colloidal fractal dimension values obtained 
from small angle x-ray scattering (SAXS) measurements \cite{Feder}. Silica colloid clusters in solution have been studied via electron microscopy, light scattering and SAXS techniques. Unlike our measurements, these experiments directly measure the bulk characteristics such as the order parameter $\beta$ in the x-ray LC scattering measurements. The \DF~values obtained from these measurements are found to be in the vicinity of {\DF${\sim}2$} for the samples with low silica content. In these experiments, fine powders of silica are compressed in order to increase 
the silica density. As in the case of our high $\rho_s$ samples, a similar increase in the calculated fractal colloid dimension has been observed at high silica densities where the fractal dimension reaches {\DF${\sim}2.5$} \cite{Feder}. 

The gels of LC+aerosil mixtures can create systems with controlled random disorder, and our experiments have demonstrated that this randomness in the bulk reflects itself at the boundary as fractal surfaces. This reveals the interconnection between the aerosil gel density in bulk $\rho_s$, order parameter $\beta$ and surface fractal dimension \DF. It remains an open question as to which other soft materials would feature such a bulk-boundary correspondence. In this work we used LCs to verify the existence of the bulk-boundary correspondence in soft matter, which can be seen as a proof of concept study. On the other hand, a research program dedicated to investigating the interplay between \DF~and the properties of other soft materials such as polymers, gels, lipids, and other biomaterials is needed. Furthermore, theoretical studies to understand the exact mechanisms that lead to bulk-boundary correspondence in randomly disordered soft matter are also missing. This line of research may open the door to engineering surfaces of polymers and biomaterials by controlling bulk randomness in order to obtain the desired surface wetting and friction properties, which mainly depend on surface fractality.

\subsection*{Acknowledgements}
M.R. was supported by TUBITAK grant no. 3001-115F315.

\subsection*{Author Contributions}
M.R. designed the experiments, and M.R. and R.S. performed the experiments. M.R. and \c{S}.\"{O} analyzed the data, interpreted the results and wrote the paper.

\bibliographystyle{naturemag}    
\bibliography{FractalLC.bib}

\vspace{1cm}

\subsection*{Methods}

In this study, we used 8CB LCs (Frinton Laboratories) without purification process as they have been used in previous x-ray and heat capacity experiments  \cite{Sungil, Germano, Leheny}.
Our hydrophilic aerosils (Evonik Corporation) were type 300 silica nanoparticles which were {$\sim7$ nm} in diameter.
The Brunauer-Emmet-Teller (BET) surface 
area for these nanoparticles is listed by the producer as 300 m$^2$g$^{-1}$. Before mixing with 8CB LC, the aerosil was dried at elevated temperatures of $T\sim$500 K under a vacuum of $\sim10^{-2}$ atm
for more than a week. The LC+aerosil mixtures were prepared according to stoichiometric ratios using the density equation $\rho_s=m_\mathrm{aerosil}/V_\mathrm{LC}$. This value can be obtained from $\rho_s^{-1}=\rho^{-1}-\rho_\mathrm{aerosil}^{-1}$ where $\rho_\mathrm{aerosil}$ is 2.2 g/cm$^{3}$. Here, $\rho$ is the ratio of the aerosil mass to the total volume of the sample, 
and $\rho_s$ is the aerosil gel density. This parameter is used to define the strength of the disorder effects created by aerosil dispersion in LC \cite{mmt,mmt1,mmt2,Salci,Sungil}. 
The aerosil nanoparticles were mixed with high purity ethanol. Each sample was sonicated at $\sim300$ K for 30 minutes. These liquid mixtures were then placed on a hot plate for the gelation process. The temperature was held in the vicinity of 8CB's isotropic phase 
temperature $\sim$310$\pm$0.5 K so that all samples were dried in the isotropic phase. 

The gelation process takes several days depending on the amount of aerosil mass in the mixture.
The gel samples prepared with the low concentration of $\rho_s\lesssim$ 0.347 g/cm$^3$ were placed on Si wafers using a spatula which was held at $\sim310$ K, the same as the drying process temperature. The samples on the Si wafers were then relaxed again at the drying temperature on the hot plate for $\sim24$ hours. For the two highest $\rho_s$ concentrations in Table I, the drying process produced small cracks on the sample surfaces. These samples were transferred to the Si surface using warm tweezers held again at $\sim307$ K. The samples were carefully held at the drying temperature during the transfer and they were relaxed after the transfer process lest the LC+aerosil gel network be externally disturbed or thermally stressed. Thus, high quality 8CB+aerosil samples were prepared carefully without any undesirable crystallization or phase separation issues.          

The AFM scans were conducted at the Nanotechnology Research Center (ITUnano) at Istanbul Technical University, a clean room facility where the temperature and
humidity were held constant at 296 K and $\sim$35\%, respectively. This temperature is just 6 K over the crystallization temperature of the pure 8CB; therefore, we avoided any unwanted accidental crystallization during the surface scans.  
The scan area for $\rho_s$=0.051, 0.078, 0.104, 0.22 and 0.347 g/cm$^3$ was $5~\mu$m$^2$, and for $\rho_s$=0.49 and 0.647 g/cm$^3$ it was $2~\mu$m$^2$.
The 3D surface pictures can be found elsewhere \cite{Salci}. 

The AFM (Nanomagnetics) was used in non-contact mode at a frequency of $\sim161$ kHz. The AFM scans were transferred to MATLAB and Mathematica using NMI Image Analyzer v1.4 software. All of the surface analyses described in the main text were performed using our own codes.

\subsection*{}

\begin{figure}
~~~~~~~~~~~~~~~~~~~~~\includegraphics[scale=0.8]{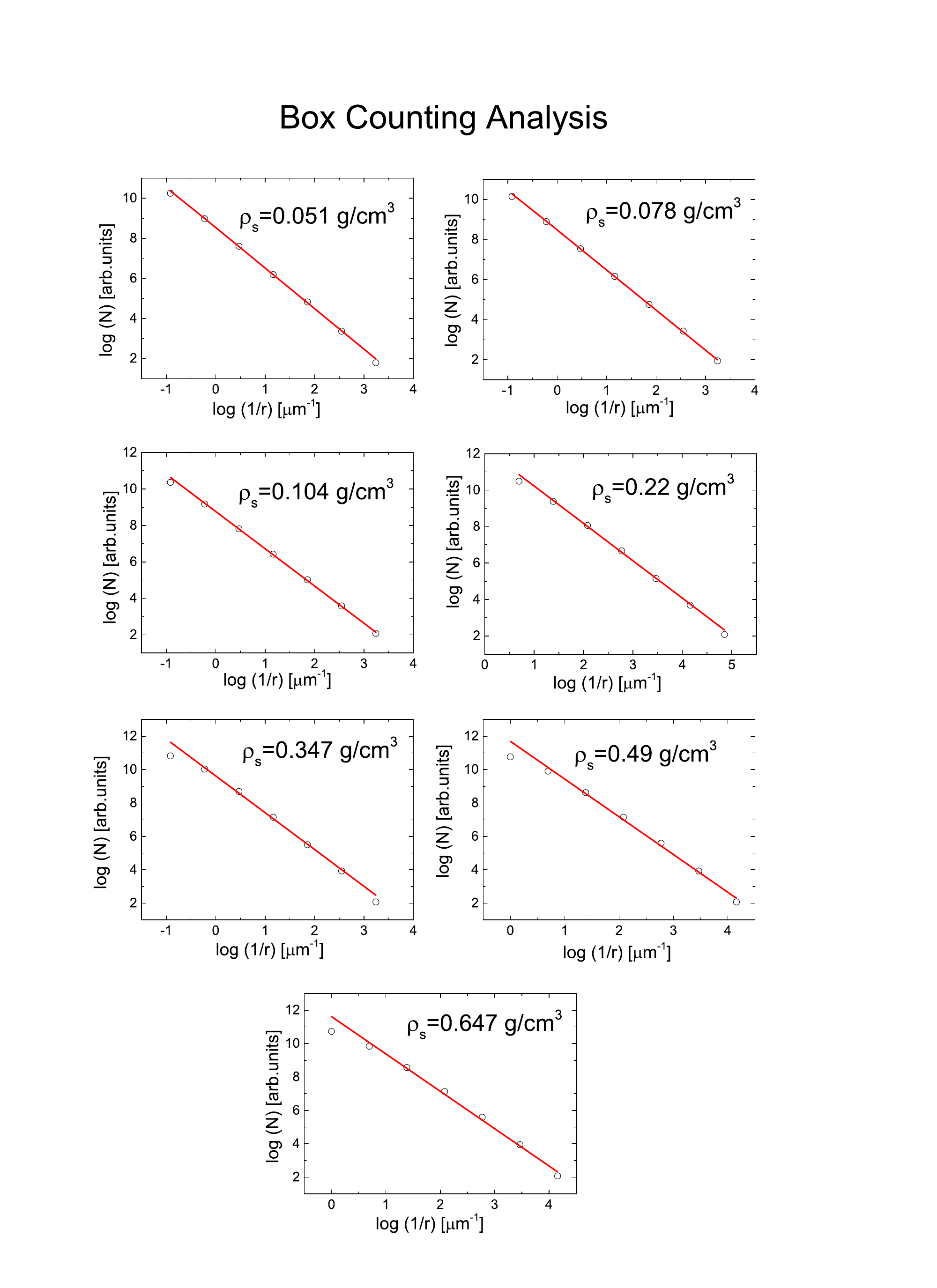}

\end{figure}

\begin{figure}
~~~~~~~~~~~~~~~~~~~~~\includegraphics[scale=0.8]{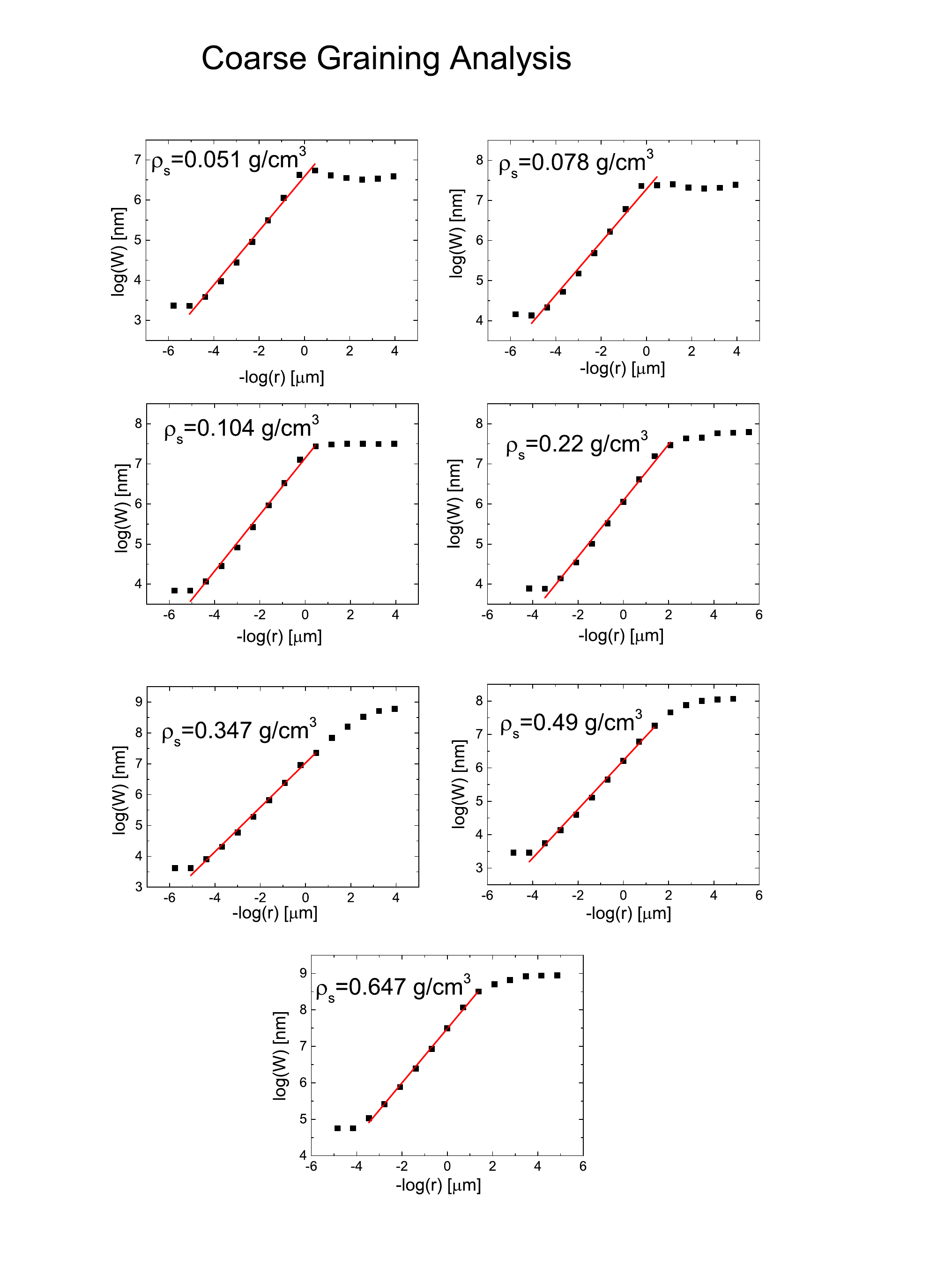}

\end{figure}

\begin{figure}
~~~~~~~~~~~~~~~~~~~~~\includegraphics[scale=0.8]{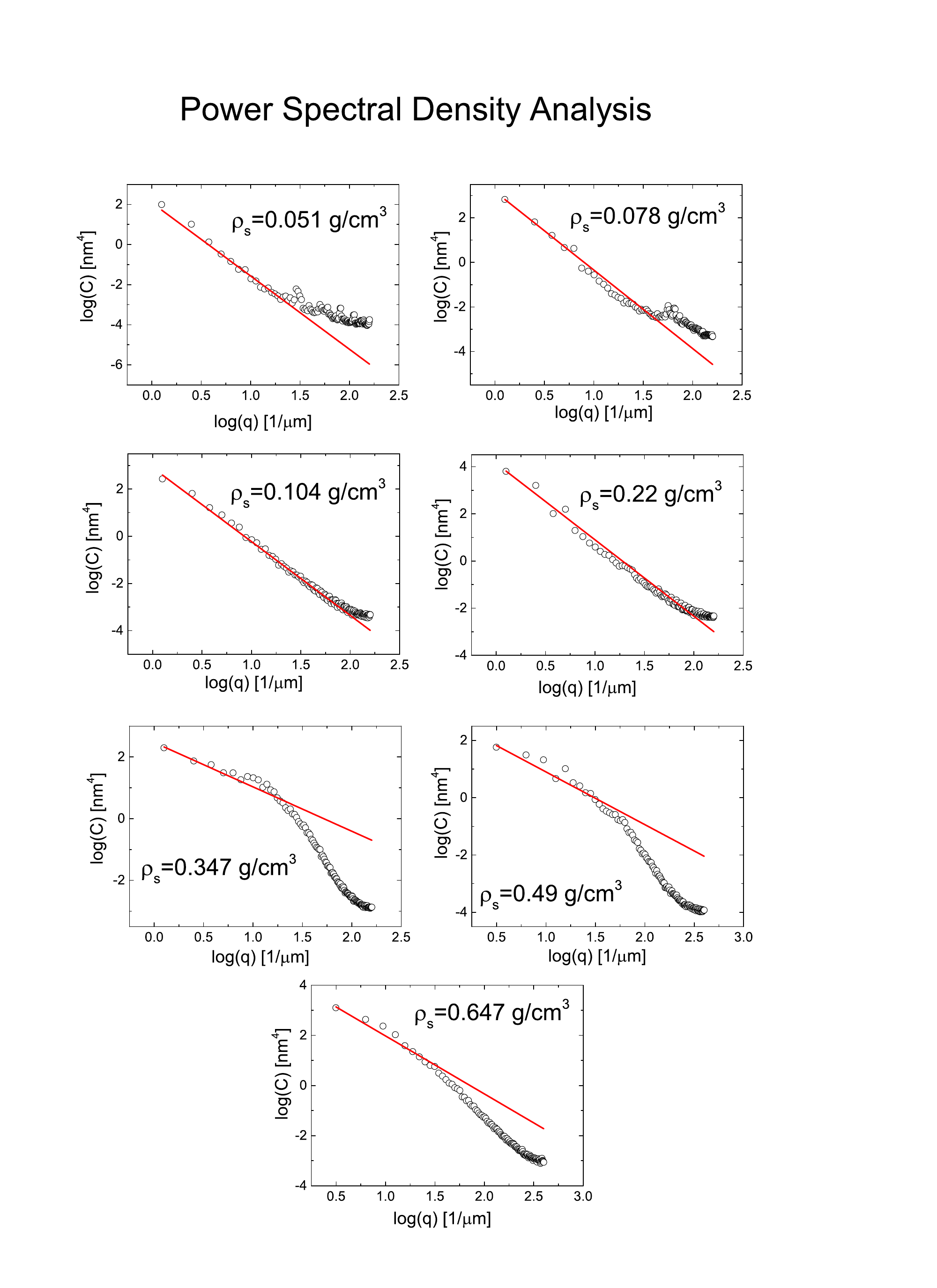}

\end{figure}

\end{document}